\documentstyle[12pt]{article}

\skewchar\fivmi='177
\skewchar\sixmi='177
\skewchar\sevmi='177
\skewchar\egtmi='177
\skewchar\ninmi='177
\skewchar\tenmi='177
\skewchar\elvmi='177
\skewchar\twlmi='177
\skewchar\frtnmi='177
\skewchar\svtnmi='177
\skewchar\twtymi='177
\def\@magscale#1{ scaled \magstep #1}
\skewchar\fivsy='60
\skewchar\sixsy='60
\skewchar\sevsy='60
\skewchar\egtsy='60
\skewchar\ninsy='60
\skewchar\tensy='60
\skewchar\elvsy='60
\skewchar\twlsy='60
\skewchar\frtnsy='60
\skewchar\svtnsy='60
\skewchar\twtysy='60

\catcode`@=11
\def\un#1{\relax\ifmmode\@@underline#1\else
        $\@@underline{\hbox{#1}}$\relax\fi}
\catcode`@=12

\def\a{\alpha}
\def\b{\beta}

\def\d{\delta}
\def\e{\epsilon}

\def\g{\gamma}

\def\l{\lambda}

\def\q{\theta}
\def\r{\rho}
\def\s{\sigma}

\def\z{\zeta}

\def\G{\Gamma}

\def\L{\Lambda}
\def\O{\Omega}

\def\dslash{\not{\hbox{\kern-2pt $\partial$}}}
\def\Dslash{\not{\hbox{\kern-4pt $D$}}}
\def\pslash{\not{\hbox{\kern-2.3pt $p$}}}
 \newtoks\slashfraction
 \slashfraction={.13}
 \def\slash#1{\setbox0\hbox{$ #1 $}
 \setbox0\hbox to \the\slashfraction\wd0{\hss \box0}/\box0 }

\font\ro=cmsy10                          
\def\kcr{{\hbox{\ro \char'170}}}                
\def\ktl{{\hbox{\ro \char'170}}}        
\def\ktr{{\hbox{\ro \char'170}}}        
\def\kbl{{\hbox{\ro \char'170}}}        
\def\kbr{{\hbox{\ro \char'170}}}        

\def\plpl{\raise-2pt\hbox{$\raise3pt\hbox{$_+$}\hskip-6.67pt\raise0.0pt
\hbox{$^+$}\hskip 0.01pt$}}
\def\mimi{\raise-2pt\hbox{$\raise3pt\hbox{$_-$}\hskip-6.67pt\raise0.0pt
\hbox{$^-$}\hskip 0.01pt$}}

\def\bo{{\raise.15ex\hbox{\large$\Box$}}}               
\def\pa{\partial}                                       
\def\de{\nabla}                                         
\def\su{\sum}                                           
\def\TH{{\raise.2ex\hbox{$\displaystyle \bigodot$}\mskip-4.7mu \llap H \;}}
\def\face{{\raise.2ex\hbox{$\displaystyle \bigodot$}\mskip-2.2mu \llap {$\ddot
        \smile$}}}                                      

\def\dvm{\raisebox{-.45ex}{\rlap{$=$}} }
\def\DM{{\scriptsize{\dvm}}~~}

\def\lin{\vrule width0.5pt height5pt depth1pt}
\def\dpx{{{ =\hskip-3.75pt{\lin}}\hskip3.75pt }}

\def\sp#1{{}^{#1}}                              
\def\sb#1{{}_{#1}}                              
\def\Tilde#1{\widetilde{#1}}                    
\def\Hat#1{\widehat{#1}}                        
\def\Bar#1{\overline{#1}}                       
\def\leftrightarrowfill{$\mathsurround=0pt \mathord\leftarrow \mkern-6mu
        \cleaders\hbox{$\mkern-2mu \mathord- \mkern-2mu$}\hfill
        \mkern-6mu \mathord\rightarrow$}
\def\dvec#1{\vbox{\ialign{##\crcr
        \leftrightarrowfill\crcr\noalign{\kern-1pt\nointerlineskip}
        $\hfil\displaystyle{#1}\hfil$\crcr}}}           

\def\fracm#1#2{\hbox{\large{${\frac{{#1}}{{#2}}}$}}}
\def\frac#1#2{{\textstyle{#1\over\vphantom2\smash{\raise.20ex
        \hbox{$\scriptstyle{#2}$}}}}}                   
\def\ha{\frac12}                                        
\def\sfrac#1#2{{\vphantom1\smash{\lower.5ex\hbox{\small$#1$}}\over
        \vphantom1\smash{\raise.4ex\hbox{\small$#2$}}}} 
\def\bfrac#1#2{{\vphantom1\smash{\lower.5ex\hbox{$#1$}}\over
        \vphantom1\smash{\raise.3ex\hbox{$#2$}}}}       
\def\afrac#1#2{{\vphantom1\smash{\lower.5ex\hbox{$#1$}}\over#2}}    

\newskip\humongous \humongous=0pt plus 1000pt minus 1000pt

\newif\ifdtup

\def\ref#1{$\sp{#1)}$}

\parskip=\medskipamount                 
\thicklines                         

\def\oldheadpic{                                
        \setlength{\unitlength}{.4mm}
        \thinlines
        \par
        \begin{picture}(349,16)
        \put(325,16){\line(1,0){4}}
        \put(330,16){\line(1,0){4}}
        \put(340,16){\line(1,0){4}}
        \put(335,0){\line(1,0){4}}
        \put(340,0){\line(1,0){4}}
        \put(345,0){\line(1,0){4}}
        \put(329,0){\line(0,1){16}}
        \put(330,0){\line(0,1){16}}
        \put(339,0){\line(0,1){16}}
        \put(340,0){\line(0,1){16}}
        \put(344,0){\line(0,1){16}}
        \put(345,0){\line(0,1){16}}
        \put(329,16){\oval(8,32)[bl]}
        \put(330,16){\oval(8,32)[br]}
        \put(339,0){\oval(8,32)[tl]}
        \put(345,0){\oval(8,32)[tr]}
        \end{picture}
        \par
        \thicklines
        \vskip.2in}
\def\oldtitle#1#2#3#4{\oldheadpic\begin{center}\vglue.5in{\large\bf #1}\\[.6in]
        {#2}\\[.1in] {\it Department of Physics and Astronomy}\\
        {\it University of Maryland, College Park, MD 20742}\\[.6in]
        Physics Publication \#{#3}\\ {#4}\\[1.5in] {\bf ABSTRACT}\\[.1in]
        \end{center} \begin{quotation}}                 
\def\oldTitle#1#2#3#4#5#6#7{\oldheadpic\begin{center} \vglue .4in
        {\large\bf #1}\\[.4in]
        {#2}\\[.1in] {\it Department of Physics and Astronomy}\\
        {\it University of Maryland, College Park, MD 20742}\\[.1in]
        {#3}\\[.1in] {\it {#4}}\\ {\it {#5}}\\[.4in]
        Physics Publication \#{#6}\\ {#7}\\[.5in] {\bf ABSTRACT}\\[.1in]
        \end{center} \begin{quotation}}                 
\def\border{                                            
        \setlength{\unitlength}{1mm}
        \newcount\xco
        \newcount\yco
        \xco=-21
        \yco=12
        \begin{picture}(140,0)
        \put(\xco,\yco){$\ktl$}
        \advance\yco by-1
        {\loop
        \put(\xco,\yco){$\kcr$}
        \advance\yco by-2
        \ifnum\yco>-240
        \repeat
        \put(\xco,\yco){$\kbl$}}
        \xco=158
        \yco=12
        \put(\xco,\yco){$\ktr$}
        \advance\yco by-1
        {\loop
        \put(\xco,\yco){$\kcr$}
        \advance\yco by-2
        \ifnum\yco>-240
        \repeat
        \put(\xco,\yco){$\kbr$}}
        \put(-20,13){\tiny University of Maryland Elementary Particle
Physics University of Maryland Elementary Particle Physics University of
Maryland Elementary Particle Physics}
        \put(-20,-241.5){\tiny University of Maryland Elementary
Particle Physics University of Maryland Elementary Particle Physics
University of Maryland Elementary Particle Physics}
        \end{picture}
        \par\vskip-8mm}
\def\bordero{                                           
        \setlength{\unitlength}{1mm}
        \newcount\xco
        \newcount\yco
        \xco=-31
        \yco=12
        \begin{picture}(140,0)
        \put(\xco,\yco){$\ktl$}
        \advance\yco by-1
        {\loop
        \put(\xco,\yco){$\kclr}
        \advance\yco by-2
        \ifnum\yco>-240
        \repeat
        \put(\xco,\yco){$\kbl$}}
        \xco=151
        \yco=12
        \put(\xco,\yco){$\ktr$}
        \advance\yco by-1
        {\loop
        \put(\xco,\yco){$\kcr$}
        \advance\yco by-2
        \ifnum\yco>-240
        \repeat
        \put(\xco,\yco){$\kbr$}}
        \put(-20,12){\ooo
bacdefghidfghghdhededbihdgdfdfhhdheidhdhebaaahjhhdahba

hgdedge
   hgfdiehhgdigicba}
        \put(-20,-241.5){\ooo
ababaighefdbfghgeahgdfgafagihdidihiidhiagfedhadbfd

ecdcdfa
   gdcbhaddhbgfchbgfdacfediacbabab}
        \end{picture}
        \par\vskip-8mm}
\def\headpic{                                           
        \indent
        \setlength{\unitlength}{.4mm}
        \thinlines
        \par
        \begin{picture}(29,16)
        \put(165,16){\line(1,0){4}}
        \put(170,16){\line(1,0){4}}
        \put(180,16){\line(1,0){4}}
        \put(175,0){\line(1,0){4}}
        \put(180,0){\line(1,0){4}}
        \put(185,0){\line(1,0){4}}
        \put(169,0){\line(0,1){16}}
        \put(170,0){\line(0,1){16}}
        \put(179,0){\line(0,1){16}}
        \put(180,0){\line(0,1){16}}
        \put(184,0){\line(0,1){16}}
        \put(185,0){\line(0,1){16}}
        \put(169,16){\oval(8,32)[bl]}
        \put(170,16){\oval(8,32)[br]}
        \put(179,0){\oval(8,32)[tl]}
        \put(185,0){\oval(8,32)[tr]}
        \end{picture}
        \par\vskip-6.5mm
        \thicklines}
\def\title#1#2#3#4{\border\headpic {\hbox to\hsize{#4 \hfill UMDEPP #3}}\par
        \begin{center} \vglue .5in {\large\bf #1}\\[.6in]
        {#2}\\[.1in] {\it Department of Physics and Astronomy}\\
        {\it University of Maryland, College Park, MD 20742}\\[1.5in]
        {\bf ABSTRACT}\\[.1in] \end{center} \begin{quotation}}  
\def\Title#1#2#3#4#5#6#7{\border\headpic
        {\hbox to\hsize{#7 \hfill UMDEPP #6}}\par
        \begin{center} \vglue .4in {\large\bf #1}\\[.4in]
        {#2}\\[.1in] {\it Department of Physics and Astronomy}\\
        {\it University of Maryland, College Park, MD 20742}\\[.1in]
        {#3}\\[.1in] {\it {#4}}\\ {\it {#5}}\\[.5in] {\bf ABSTRACT}\\[.1in]
        \end{center} \begin{quotation}}                 
\def\endtitle{\end{quotation}\newpage}                  

\def\sect#1{\bigskip\medskip \goodbreak \noindent{\bf {#1}} \nobreak \medskip}

\begin{document}


\def\dvm{\raisebox{-.45ex}{\rlap{$=$}} }
\def\DM{{\scriptsize{\dvm}}~~}

\def\lin{\vrule width0.5pt height5pt depth1pt}
\def\dpx{{{ =\hskip-3.75pt{\lin}}\hskip3.75pt }}
\def\fracm#1#2{\hbox{\large{${\frac{{#1}}{{#2}}}$}}}

\border\headpic {\hbox to\hsize{December 1994 \hfill UMDEPP 95-75}}\par
\begin{center}
\vglue .4in
{\large\bf Vector Multiplets
and the \\ Phases of N = 2 Theories in 2D \\
Through the Looking Glass \footnote{Research supported
by NSF grant \# PHY-93-41926}
${}^,$ \footnote {Supported in part by NATO Grant CRG-93-0789}
}\\[.2in]
S. James Gates, Jr.\footnote{gates@umdhep.umd.edu} \\[.1in]
{\it Department of Physics\\
University of Maryland at College Park\\
College Park, MD 20742-4111, USA}
\\[2.5in]

{\bf ABSTRACT}\\[.1in]
\end{center}
\begin{quotation}

We extend Witten's discussion of actions related to the Landau-Ginzburg
description of Calabi-Yau hypersurfaces in weighted projective spaces
to cover the mirror class of models that include twisted chiral matter
multiplets and a newly discovered 2D, N = 2 twisted vector multiplet.
Certain integrability obstructions are observed that constrain the most
general constructions containing both matter and twisted matter
simultaneously.  It is conjectured that knot invariants will ultimately
play a role in describing the most general such model.
\endtitle

\sect{I. Introduction}

Over a decade ago, it was noted that 2D supersymmetry representations
included some unusual (from the view of 4D) representations, the
twisted chiral multiplets \cite{SJG1,SJG2}. Our discovery began
precisely with the problem of the toroidal compactification of
4D, N = 2 vector multiplets to the 2D, N = 4 and N = 2 theories.
The output of those investigations led to the {\underline {first}}
appearance in the literature of special K\" ahler geometries,
K\" ahlerian N = 2 vector multiplets\footnote{Recently \cite{SW},
4D K\" ahlerian vector multiplet theories have been used to
make progress in \newline ${~~~~~}$ understanding a number of issues.},
the use of duality as a tool for finding new supersymmetric representations
and the appearance of torsion in 2D nonlinear sigma-models. All
of these topics have reappeared with a vengeance in compactified
heterotic and superstring theory where they play important roles.

Although twisted chiral multiplets were discovered prior to
the advent of compactified heterotic string theory, nevertheless
they encode the local part of the ``mirror symmetry transformation.''
They allow Lagrangian field theories to represent both (c,c) and
(a,c) rings in the language of superconformal field theory. The
local part of mirror symmetry, in its simplest form,  is just
the statement that given one action written in terms of chiral
matter scalars, there exist another theory where all of
the chiral scalars are replaced by twisted chiral scalars.

In more recent times Witten \cite{WTN}, has used similar techniques
applied to supersymmetric world sheet actions to show how the
Calabi-Yau hypersurfaces in weighted projective spaces occur as
a phase of 2D, N = 2 supersymmetric Landau-Ginzburg type models.
The matter field sectors of his models included only chiral
scalar multiplets. One of the questions left open in this
investigation was how the extension of such techniques might apply
to models where the matter fields included twisted chiral
multiplets.

It is the purpose of this paper to research this open question. In
the following we will gaze ``through the looking glass'' at these
models. In other words, we will perform a mirror transformation on
the models described in reference one to obtain their ``mirror
images.''   This will be a new class of models where the matter
multiplets are described by twisted chiral superfields.  In order
to achieve this, it will be necessary to introduce a previously
unknown representation of 2D, N = 2 supersymmetry; the twisted
vector multiplet.  Since the ordinary vector multiplet is described
by a twisted chiral scalar field strength, the twisted vector multiplet
is described by an ordinary chiral scalar field strength.

\sect{II. The Mirror Image of the 2D, N = 2 Vector Multiplet}

We first describe the usual 2D, N = 2 vector multiplet (VM-I) in 2D,
N = 2 superspace. For this purpose we introduce a superspace Yang-Mills
covariant derivative $\nabla_A \equiv D_A ~+~ i g \G_A t$ (in this expression
$t$ denotes a U(1) Lie algebra generator) that has a
supercommutator algebra given by,
$$
[\de \sb{\a} , \de \sb{\b} \}  ~=~ 0
{}~~~, ~~~
[\de \sb{\a} , \Bar \de \sb{\b} \} ~=~  i 2 (\g \sp{c}) \sb{\a \b} \de
\sb{c} ~+~ ~ 2 g [~ C_{\a \b} S ~-~ i (\g^3 )_{\a \b} P ~] t ~~~,
$$
\begin{equation}  {~~~~~~~~}
[\de \sb{\a} , \de \sb{b} \} ~=~ g (\g_b )_{\a}{}^{\b} {\Bar W}_{\b} t
{}~~~, ~~~~~
[\de \sb{a} , \de \sb{b} \} ~=~   - i g \e_{a b} {\cal W} t ~~~, ~~~
\end{equation}
and
$$ \nabla_{\a} S ~=~ - i{\Bar W}_{\a} ~~~,~~~ \nabla_{\a} P ~=~ -
(\g^3)_{\a}{}^{\b}  {\Bar W}_{\b} ~~~,~~~ \nabla_{\a} {\Bar W}_{\b}
{}~=~ 0  ~~~,  ~~~\nabla_{\a} {\rm d} ~=~ (\g^c )_{\a}{}^{\b}  \de \sb{c}
{\Bar W}_{\b} ~~~, $$
\begin{equation}
{\nabla}_{\a}  W_{\b} ~=~ i  C_{\a \b} {\rm d} ~-~
(\g^3)_{\a \b} {\cal W} ~+~ (\g^a)_{\a \b} ( \nabla_a S)
{}~-~ i (\g^3 \g^a)_{\a \b} ( \nabla_a P) ~~~.
\end{equation}
This is the vector multiplet that comes down via dimensional reduction
from higher dimensions.  The standard kinetic action for a U(1) gauge group
for this multiplet forms a supersymmetric invariant,
$${\cal S}_{\rm{VM-I}} ~=~ \int d^2 \s  d^2 {\bar \z} d^2 \z  ~[~- \frac 1{4}
S^2 ~] ~=~ \int d^2 \s  d^2 {\bar \z} d^2 \z ~[~- \frac 1{4} P^2 ~] ~=~
{~~~~~~~~~~~~~~~~~~~~~~~~~~} $$
\begin{equation}
{~~~~~~~~~} = \int d^2 \s ~ [ - \fracm 14
({F}_{a b}(A))^2 ~+~ \fracm 12 ( \nabla_a S)^2 ~+~ \fracm 12
( \nabla_a P)^2 ~-~ i {\bar \l}_{\a} (\g \sp{c}) \sp{\a \b}
\nabla_c {\l}_{\b} ~+~ \fracm 12 {\rm d}^2 ~] ~~~,
\end{equation}
where we have used the facts that ${\cal W} | =\frac 12
 \e^{a b} F_{a b} (A)$ and $W_{\a} | = \l_{\a}$. (It is also of
interest to note that we utilize a definition $\int  d^2 {\bar \z} d^2 \z
\equiv \frac 18 [ {\Bar \nabla}^{\a} {\Bar \nabla}_{\a} { \nabla}^{\b}
{\nabla}_{\b} + {\nabla}^{\a} {\nabla}_{\a} {\Bar \nabla}^{\b} {\Bar
\nabla}_{\b} ]$.)   One of the interesting properties of the vector
multiplet is that it can be coupled to chiral superfields but cannot
be coupled to twisted chiral superfields.  The proof of this statement
can be seen as follows. We can use the following notations for $\nabla_+$,
$\nabla_-$, ${\Bar \nabla}_+$, and ${\Bar \nabla}_-$.
$$
\nabla_+ ~\equiv~ \frac 12 (~1 ~+~ \g^3 )_{\a} {}^{\b} \nabla_{\b} ~~~,~~~
\nabla_+ ~\equiv~ \frac 12 (~1 ~-~ \g^3 )_{\a} {}^{\b} \nabla_{\b} ~~~,~~~
$$
\begin{equation}
{\Bar \nabla}_+ ~\equiv~ \frac 12 (~1 ~+~ \g^3 )_{\a} {}^{\b} {\Bar
\nabla}_{\b} ~~~,~~~
{\Bar \nabla}_- ~\equiv~ \frac 12 (~1 ~-~ \g^3 )_{\a} {}^{\b} {\Bar
\nabla}_{\b}  ~~~.
\end{equation}
Defining the linear combination ${\Psi} \equiv S ~+~ i P$, we see
that $\nabla_+ {\bar \Psi} = 0$ and ${\Bar \nabla}_- {\bar \Psi} = 0$.
This identifies $\Psi$ as a twisted chiral superfield. We now introduce
a matter scalar multiplet ${\bar \chi}$ that is also a twisted chiral
superfield. We thus have,
\begin{equation}
\nabla_+ {\bar \chi} = 0 \to {\Bar \nabla}_- \nabla_+ {\bar \chi} = 0 ~~~,
\end{equation}
\begin{equation}
{\bar \nabla}_- {\Bar \chi} = 0 \to  \nabla_+ {\Bar \nabla}_- {\bar \chi} = 0
{}~~~,
\end{equation}
and adding these two results together yields
\begin{equation}
 i 2 g {\bar \Psi} [~ t , {\bar \chi} ] ~=~ 0 ~~~,
\end{equation}
after taking the appropriate chiral projections of the second
result in equation (1.). This is the standard integrability-type
argument that often occurs in superspace.  The only reasonable
way to satisfy this condition is to demand that $\chi$ lie in the
trivial representation of the gauge group of the vector multiplet
or equivalently the matter field $\chi$ does not carry a
charge to which the gauge field couples.

We now come to the ``mirror image'' of the usual vector multiplet.
Below we will see that there is actually a second such multiplet!
Initially this may be extremely surprising to the reader. This should
not be the case.   Mirror symmetry is apparently a fundamental part
of 2D, N = 2 supersymmetry.  Mirror symmetry is connected with
the definition of a 2D parity operator.  For example, the most
important difference between a 2D chiral multiplet and a 2D
twisted chiral multiplet is that the former includes two scalar
spin-0 fields in its spectrum while the latter includes one scalar
and one pseudoscalar in its spectrum.  We have a precedent
for this behavior in 3D, N = 4 theories \cite{BrkGt} as well as 2D,
N = 2 supergravity theories \cite{GLO}.

The second vector multiplet (VM-II) can be introduced in the following
manner.  For this purpose we introduce a superspace Yang-Mills
covariant derivative $\nabla_A \equiv D_A ~+~ ig' {\Tilde \G}_A t'$
that has a supercommutator algebra given by,
$$
[\de \sb{\a} , \de \sb{\b} \}  ~=~ i 4 g' (\g^3 )_{\a \b} {\bar {\cal P}}
t' ~~~, ~~~
[\de \sb{\a} , \Bar \de \sb{\b} \} ~=~  i 2 (\g \sp{c}) \sb{\a \b} \de
\sb{c}  ~~~,  ~~~
$$
\begin{equation}
[\de \sb{\a} , \de \sb{b} \} ~=~ - g' (\g^3 \g_b )_{\a}{}^{\b} {\bar \O}_{\b}
t' ~~~, ~~~
[\de \sb{a} , \de \sb{b} \} ~=~
  - i g' \e_{a b} { {\cal U}} t' ~~~, ~~~
\end{equation}
and
$$ { ~~~~~~~~ } \nabla_{\a} {\bar {\cal P}} ~=~ 0 ~~~,~~~ {\Bar \nabla}_{\a}
{\bar {\cal P}} ~=~ {\bar \O}_{\b}
  ~~~,~~~ \nabla_{\a} {\bar \O}_{\b}
{}~=~  i 2 (\g^a)_{\a \b} ( \nabla_a {\bar {\cal P}})  ~~~, $$
\begin{equation}
{\nabla}_{\a} { \O}_{\b} ~=~  C_{\a \b} ~[ ~
 { {\cal U}} ~+~ i {\tilde {\rm d}} ~] ~~~,~~~ \nabla_{\a} {\tilde
{\rm d}} ~=~ (\g^c )_{\a}{}^{\b}  \de \sb{c}  {\bar \O}_{\b} ~~~.
\end{equation}
This vector multiplet cannot be obtained via dimensional reduction
from higher dimensions. The kinetic action for a U(1) gauge group
for the multiplet is the mirror image of the standard one above.
$${\cal S}_{\rm{VM-II}} ~=~ \int d^2 \s  d^2 {\bar \z} d^2 \z ~[~ \frac 1{2}
{\bar {\cal P}} {\cal P} ~] ~=~ {~~~~~~~~~~~~~~~~~~~~~~~~~~~~~~~~~~~~~~~~~}
$$
\begin{equation}
{~~~~~~~~~~~~~~~~~~~} = \int d^2 \s ~ [ - \fracm 14 ({F}_{a b}(B))^2 ~+~ 2
| \nabla_a {\cal P} |^2 ~-~ i {\bar \r}_{\a} (\g \sp{c})
\sp{\a \b} \nabla_c {\r}_{\b} ~+~ \fracm 12 {\tilde {\rm d}}^2 ~] ~~~,
\end{equation}
where ${{\cal U}} | = \frac 12 \e^{a b} F_{a b} (B)$
and ${\O}_{\a} | = \r_{\a}$.

It has the mirror image property to that of the VM-I theory in that
it can be coupled to twisted chiral superfields but cannot be coupled to
chiral superfields. The argument is just the mirror image of that
given in equations (4-6). We note that the definition of a chiral
matter scalar ${\bar \Phi}$ implies the following integrability argument.
\begin{equation}
\nabla_{\a} {\bar \Phi} = 0 \to \nabla_{\b} \nabla_{\a} {\bar \Phi} = 0
\to  [~ \nabla_{\b} , \nabla_{\a} ~ \} {\bar \Phi} = 0 \to ~
i4 g' ~(\g^3)_{\a \b} {\bar {\cal P}} [~ t' , {\bar \Phi} ~] = 0 ~~~.
\end{equation}

We end this section by considering the unconstrained prepotential
formulation of the twisted vector multiplet.  For simplicity
we will only consider abelian theories. It is well known that a
chiral (anti-chiral) superfield can be obtained from a general
complex superfield $U$ via the equations
\begin{equation} {\Phi} ~ \equiv ~  \frac 12 C^{\a \b}
{\bar D}_{\a} {\bar D}_{\b} U  ~~~,  ~~~
{\bar \Phi} ~ \equiv ~ \frac 12 C^{\a \b} D_{\a}
{D}_{\b} {\Bar U} ~~~.
\end{equation}
What is less well known is that similar equations apply to twisted
chiral scalar multiplet also.
$$
S ~\equiv~ \frac 14 [~ C^{\a \b} D_{\a} {\bar D}_{\b} (~ U ~+~ {\Bar U} ~)
{}~+~ (\g^3 )^{\a \b}  D_{\a} {\bar D}_{\b} (~ U ~-~ {\Bar U} ~) ~] ~~~,
{~~} $$
$$
P ~\equiv~ - i \frac 14 [~ (\g^3 )^{\a \b}  D_{\a} {\bar D}_{\b} (~ U ~+~
{\Bar U} ~) ~+~ C^{\a \b} D_{\a} {\bar D}_{\b} (~ U ~-~ {\Bar U} ~) ~]
 ~~,
$$
\begin{equation}
\chi ~=~ \frac 12 (~1 ~+~ \g^3 ~)^{\a \b}  D_{\a} {\bar D}_{\b} U ~~~,~~~
{\bar \chi} ~=~ \frac 12 (~1 ~-~ \g^3 ~)^{\a \b}  D_{\a} {\bar D}_{\b}
{\Bar U} ~~~.
\end{equation}
These last two equations are extremely useful because they determine
the structure of the superpropagators for twisted chiral superfields.
Finally we come to the vector multiplets.

For the usual vector multiplet the covariant derivative ($\de_A \equiv
D_A + i g \G_A t$) can be explicitly expressed in terms of a prepotential
superfield $V$ that is the fundamental gauge superfield of any SUSY YM-type
theory. In particular the components of the superconnection are defined
by,
$$
\G_{\a} ~=~ i D_{\a} V ~~~,~~~{\Bar \G}_{\a} ~=~ - i {\bar D}_{\a} V
{}~~~, ~~~ $$
\begin{equation} { ~~~}
\G_{a} ~=~  \frac 14 (\g_a)^{\a \b} ~[~ ({D}_{\a} {\bar D}_{\b}
 ~-~ {\bar D}_{\b} {D}_{\a} )V ~] ~~~,
\end{equation}
and their gauge variations follow from that of the prepotential
$\d_G V = - i ( \L - {\bar \L} ) $ where $\L$ is a chiral superfield.
For the twisted vector multiplet the covariant derivative (${\Tilde \de}_A
\equiv D_A + i g' {\Tilde \G}_A t$) can also explicitly expressed in terms
of a prepotential superfield ${\tilde V}$ that is a fundamental gauge
superfield.   In particular the components of the superconnection now are
defined by,
$$
{\Tilde \G}_{\a} ~=~ i (\g^3)_{\a}{}^{\b} D_{\b} {\Tilde V} ~~~,~~~ {\Tilde
{\Bar \G}}_{\a} ~=~ - i (\g^3)_{\a}{}^{\b} {\bar D}_{\b} {\Tilde V}
{}~~~, ~~~
$$
\begin{equation}
{\Tilde \G}_{a} ~=~  \frac 14 (\g^3 \g_a)^{\a \b} ~[~ ({D}_{\a} {\bar D}_{\b}
 ~-~ {\bar D}_{\b} {D}_{\a} ) {\Tilde V} ~] ~~~,  {~~~~~~~}
\end{equation}
and their gauge variations follow from that of the prepotential
$\d_G {\Tilde V} = - i ( {\Tilde \L} - {\Tilde {\Bar \L}} ) $ where $
{\tilde \L}$ is a twisted chiral superfield.

A final point of interest to note is that these superspace results allow a
simple generalization of a well known result for 2D gauge fields.  Any
such field ($v_a$) has a natural decomposition of the form $v_a = \pa_a \l
+ \e_a {}^b \pa_b \tilde \l $ plus a harmonic piece. This same
result holds for a supergauge prepotential $V$ in the form $V =
- i ( {\L} - {\Bar \L} ) - i ( {\Tilde \L} - {\Tilde {\Bar \L}} )$ plus
a superharmonic piece.

\sect{III.  CY-LG Model Actions Through the Looking Glass}

The most general action for chiral matter fields ($\Phi$), twisted
chiral matter ($\chi$), vector multiplets ($\Psi$) and twisted vector
multiplets (${\cal P}$) takes the form of a typical N = 2 nonlinear
$\s$-model with superpotential and twisted superpotential terms
form,
\begin{equation}
{\cal S}_{\rm K} ( \Phi, \chi: \Psi , {\cal P} ) ~=~ {\cal S}_{\rm K} ~+~
{\cal S}_{\rm W} ~+~ {\cal S}_{\Tilde {\rm W}} ~~~  ,
\end{equation}
where
\begin{equation}
{\cal S}_{\rm K} ~=~ \int d^2 \s  d^2 {\bar \z} d^2 \z ~{\rm K}( \Phi, \chi:
\Psi , {\cal P} ) ~~~,
\end{equation}
\begin{equation}
{\cal S}_{\rm W} ~=~ \int d^2 \s d^2 \z  ~{\rm W}( \Phi: {\cal P} ) ~+~
{\rm h.} {\rm c.} ~~~,
\end{equation}
\begin{equation}
{\cal S}_{\Tilde {\rm W}} ~=~ \int d^2 \s d {\bar \z}^- d \z^+  ~{\Tilde
{\rm W}} (\chi: \Psi ) ~+~ {\rm h.} {\rm c.} ~~~.
\end{equation}
The work of Witten \cite{WTN} specialized to the case ${\cal S}_{\rm K}
( \Phi, 0: \Psi , 0 )$ and the simple mirror reflection of this
would be ${\cal S}_{\rm K}  ( 0 , \chi: 0 , {\cal P} )$.   For example,
the (local part of) the mirror transformation acting on the K\" ahler-like
potential term consists of the operations
\begin{equation}
\Phi \to \frac 12 \chi ~~~,~~~ \chi \to  2 \Phi ~~~,~~~ {\cal P} \to
\frac 12 \Psi ~~~,~~~ \Psi \to 2 {\cal P} ~~~,~~~ {\rm K} \to - {\rm K}
{}~~~.
\end{equation}

The whole philosophy of the CY-LG models is to restrict the class of
kinetic energy terms to be flat. So the K\" ahler-like potential ($
{\rm K} $) takes the form of being purely quadratic in the superfields.
Thus, we have for the most general K\" ahler-like potential
\begin{equation}
{\rm K} ( \Phi, \chi: \Psi , {\cal P} ) ~=~ - \frac 18 {\bar {\Psi}}
{\Psi}  ~+~ \frac 12 {\bar {\cal P}} {\cal P} ~+~ \frac 12 {\bar \Phi}
{\Phi} ~-~ \frac 18 {\bar {\chi}} {\chi} ~~~  .
\end{equation}
It is important that we say a few words about some of the terms in
the potential above.  There is the interesting change of sign
in comparing the the kinetic energy of a chiral multiplet and
a twisted chiral multiplet.  This sign difference is no accident.
If we think of the derivatives with respect to the chiral superfields
as providing the natural basis of a tangent bundle. Then the
derivatives with respect to the twisted chiral superfields provide
the basis for the cotangent bundle. The curvature of a manifold
taken with respect to the two different bases differs by a sign.
The tangent cotangent interpretation is inherent in the presence
of mirror symmetry.  Also it may look as though we have {\underline
{not}} coupled the matter multiplets to the vector multiplets. In fact,
we have. The matter multiplets in this expression have their chiral
and twisted chirality conditions defined with respect to the U(1) $\times$
U'(1) supercovariant derivative $\nabla_A \equiv D_A ~+~ ig {\G}_A t ~+~
ig'{\Tilde \G}_A t'$  not the ``bare'' supercovariant derivatives. This
automatically insures minimal coupling.  The U(1) and U'(1) generators
are still restricted to satisfy equations (7.) and (11.). Additionally,
we have $[ t , {\Phi} ] = i{\rm Q} {\Phi}$ and $[t' , {\chi} ] = i{\rm Q}'
{\chi}$ for charges ${\rm Q}$ and ${\rm Q}'$.

This leaves the most interesting part of the actions to reside in the
superpotential and twisted superpotential terms. Expanding out the action
in (18) in terms of components yields,
$$
{\cal S}_{\rm W} ~=~ \int d^2 \s ~\frac 12 \{ ~[~ {{\rm W}''} \psi^{\a}
\psi_{\a} ~+~ 2 {{\rm W}'} F ~+~ {\ddot {\rm W}} \r^{\a} \r_{\a}
{}~+~ 2  {\dot {\rm W}}' \r^{\a} \psi_{\a}  ~+~ {\rm h.} {\rm c.} ~]
{~~~~~~} $$
\begin{equation}
{~~~~~~~~~~~~~~~~~}
{}~-~ ( {\dot {\rm W}} ~+~ {\dot {\rm W}}^* ) \e^{a b} F_{a b} (B)
{}~-~ 2 i ({\dot {\rm W}} ~-~ {\dot {\rm W}}^* ) {\tilde {\rm d}} ~
 ~\} ~~~.
\end{equation}
where ($A$, $\psi_{\a}$, $F$) are the components of the chiral
multiplet ($\Phi$) defined by $\Phi | = A$, $\nabla_{\a} \Phi | = \psi_{\a}$
and $\frac 12 \nabla^{\a} \nabla_{\a} \Phi | = F$. We have also used the
notations
\begin{equation}
{\rm W} ~=~ {\rm W} (A : {\cal P}) ~~~,~~~ {\rm W}' \equiv {{\pa  {\rm W}
\over \pa {\Phi}}} ~~~,~~~ {\dot {\rm W}} \equiv {{\pa  {\rm W}  \over
\pa {{\cal P}}}} ~~~.
\end{equation}
In a similar fashion equation (20) yields
$$
{\cal S}_{\Tilde {\rm W}} ~=~ \int d^2 \s ~\{ ~ [~ 4 {{\Tilde {\rm W}}''}
{\bar \varphi}_{-} \varphi_{+} ~+~ 2 {{\Tilde {\rm W}}'} h ~+~  4 {\ddot
{\Tilde {\rm W}}} {\bar \l}_{-} \l_{+} ~+~ 4 {\dot {\Tilde {\rm W}'}}
( \l_{-} {\bar \varphi}_{+} ~+~ {\varphi}_{-} {\bar \l}_{+} )  ~+~ {\rm h.}
{\rm c.} ~]
$$
\begin{equation}
{~~~}
{}~+~ ( {\dot {\Tilde {\rm W}}} ~+~ {\dot {\Tilde {\rm W}^*}} )
\e^{a b} F_{a b} (A) + i 2 ( {\dot {\Tilde {\rm W}}} ~-~ {\dot {\Tilde {\rm
W}^*}}
) { {\rm d}} ~\} ~~~.
\end{equation}
where ($a$, $\varphi_{\a}$, $h$) are the components of the twisted chiral
multiplet ($\chi$) defined by $\chi | = a$, $\nabla_+ \chi | = - i 2{\bar
\varphi}_+$,
${\Bar \nabla}_- \chi | = i 2 {\varphi}_-$ and ${\Bar \nabla}_- \nabla_+ \chi |
= 2 h$.
Acting on ${\tilde {\rm W}}$ we use the notations
\begin{equation}
{\Tilde {\rm W}} = {\Tilde {\rm W}} ( a : \Psi) ~~~,~~~ {\Tilde {\rm W}}'
 \equiv {{\pa  {\Tilde {\rm W}}
\over \pa {a}}} ~~~,~~~ {\dot {\Tilde {\rm W}}} \equiv {{\pa  {\Tilde
{\rm W}} \over \pa {{\Psi}}}} ~~~.
\end{equation}

It is here that the manifestation of having all the multiplets and
their full compliment of mirror images produces something new. In
particular, there is a possibility to introduce non-minimal couplings
between chiral matter and twisted vector multiplets as well as the
mirror image coupling between twisted chiral matter and vector multiplets.
These new couplings have an exceedingly interesting interpretation if we
view them as arising from the dimensional reductions from a 3D model.
Namely, we see that the presence of the most general terms in the
two types of superpotentials lead to 2D, N = 2 Chern-Simons terms
making their appearance! In (22) the quantity $( {\dot {\rm W}} + {\dot
{\rm W}}^* )$ plays the role of the ``third'' component of the gauge
field in a 3D CS action and a similar role can be seen for $( {\dot {\Tilde
{\rm W}}} + {\dot {\Tilde {\rm W}^*}} )$ in (24). It is also obvious now
that the parameter $t$ introduced by Witten can be interpreted as the
the vacuum value (moduli-like parameter) of the complex ``third'' gauge
component $\dot {\Tilde {\rm W}}$.

In the nonlinear $\s$-model formulation of ref. \cite{SJG2}, it was shown
that the introduction of torsion required that the mixed derivative of the
K\" ahler-like potential taken with respect to one chiral superfield and one
twisted chiral superfield should be non-vanishing. This suggests that within
the context of these CY-LG models, the introduction of torsion is dependent
upon the either the chiral superpotential or twisted chiral superpotential
depending upon twisted vector or vector multiplets, respectively.

\sect{IV.  Summary and Conclusion}

As we have demonstrated, the extension of the CY Landau-Ginzburg
models, as first proposed by Witten, to cover a larger class of
models that contain both chiral and twisted chiral matter is
possible.  This larger class of actions is characterized by formula
(16.) containing a new class of terms over and above those that one
would expect as simple mirror reflections. These are the non-minimal
couplings of the VM-I multiplet to twisted chiral matter and their mirror
reflections, i.e. the 2D Chern-Simons terms.  The algebraic-geometrical
significance of these new interactions is not completely clear at this
time. But the interpretation of the new terms as the result of dimensional
reduction from the 3D Chern-Simons action suggests the possibility
of intersection polynomials \cite{WTN2} playing a powerful and
previously unsuspected role.
$${~~~} $$

$${~~~} $$

\noindent {\bf{Acknowledgment} }
\indent \newline
S.~J.~G. wishes to acknowledge conversations with R.~Brooks, T.~Hubsch,
H.~Nishino and V.~G.~J.~Rodgers that contributed to the completion
of this investigation.

\newpage

\noindent{{\bf {APPENDIX A: A Comment on VM + SG Theories }}}

In the work of ref.~\cite{WTN}, there was never an introduction of local
supersymmetry into the structure of the models. However, in a later
work \cite{NSH}, this extension was studied. In this brief appendix,
we will comment upon a general aspect that seems not to have been
realized previously with regard to the introduction of spin-1 gauge
fields in the presence of N $\ge$ 1 supergravity.  Below, we will
show that generically in such theories, the removal of the supergravity
spin-0 auxiliary fields by their algebraic equations of motion
strongly restricts the appearance of the usual spin-1 field strength in
any such local theory!

We shall show this effect within the context of 2D, N = 1 SG coupled to
matter. Out starting point is the covariant description of SG + SUSY
YM (abelian) theory.
$$
[ ~ \nabla_{\a} ~,~  \nabla_{\b} ~ \} ~= ~ i 2 (\g^a)_{\a \b}
\nabla_{a} ~+~  2 (\g^3)_{\a \b} [~{\rm R} {\cal M} ~+~ i g P t~]
{}~~, {~~~~~~~~~~~~~~~~~~~~~}
\eqno(A.1) $$
$$
[ ~ \nabla_{\a} ~,~  \nabla_{b} ~ \} ~= ~ i  [ ~ \ha {\rm R}
(\g_b)_{\a}{}^{\b} \nabla_{\b} ~+~ (\g^3 \g_b)_{\a}{}^{\b}
( \nabla_{\b}{\rm R} ) {\cal M} ~-~ i g (\g_b)_{\a}{}^{\b} W_{\b}~ ] ~~,
{~~~}
\eqno(A.2) $$
$$
[ ~ \nabla_{a} ~,~  \nabla_{b} ~ \} ~= ~ - \e_{a b} ~[ ~ \ha
( \nabla^{\a} {\rm R} ) (\g^3)_{\a}{}^{\b} \nabla_{\b} ~-~
(\nabla^2 {\rm R} ~-~  {\rm R}^2 ) {\cal M} ~+~ i g {\cal W} t~ ] ~~,
\eqno(A.3) $$
The quantities ($P$, $W_{\a}$, ${\cal W}$) are the components of a
2D, N = 1 vector multiplet (where ${\cal W}| =  \frac 12 \e^{a b} F_{a b}
(A)$). These must satisfy the relations
$$ \nabla_{\a} P ~=~ (\g^3)_{\a}{}^{\b} W_{\b} ~~~,~~~ \nabla^2 P ~=~
- {\cal W} ~+~ {\rm R} P ~~~. \eqno(A.4)$$
so that the Bianchi identities on the vector multiplet are also satisfied.
Our goal is to show that any Lagrangian of the form,
$$
{\cal L} ~=~ {\rm E}^{-1} \left [ ~ \frac 14 C^{\a \b}  (\nabla_{\a} P )
(\nabla_{\b} P )  ~+~ \frac 14 C^{\a \b} { \su\limits_{i = 1}^{M} }
({\nabla}_{\a} \Phi_i )  ({\nabla}_{\b} \Phi_i )  ~+~
{\rm U} (\Phi_i : P ) ~ \right ]  ~~~,
\eqno(A.5)$$
has the property that the elimination of the supergravity auxiliary
spin-0 field by its algebraic equation of motion always leads to the
unexpected result of an action that is at most linear in $F_{a b} (A)$!
The proof is very direct. In order to find the component expression
that follows from (A.5) we note the following useful identities.
$$
\int d^2 \s d^2 \q {\rm E}^{-1} {\cal L} = \int d^2 \s {\rm e}^{-1}
[ \nabla^2 ~-~ i \psi_a {}^{\b} (\g^a)_{\b}{}^{\a} \nabla_{\a}
{}~-~ {\rm B} ~+~  \e^{b c} \psi_b {}^{\a} (\g^3 )_{\a \b} \psi_c {}^{\b}
] {\cal L} | ~~~,
\eqno(A.6)$$
$$ \nabla_{\a} \nabla_{\b} ~=~ i  (\g^a)_{\a \b}
\nabla_{a} ~+~  (\g^3)_{\a \b} {\rm R} {\cal M} ~+~ C_{\b \a }
\nabla^2 ~~~, ~~~~~~~~~~~~~~~~~~~~~~~~~~~~~~~~~~~~\eqno(A.7)$$
$$
\nabla^2 \nabla_{\a} ~=~ - i (\g^a)_{\a}{}^{\b} \nabla_{a} \nabla_{\b}
{}~+~ {\rm R} [ ~\nabla_{\a} ~-~ (\g^3)_{\a}{}^{\b} {\cal M} \nabla_{\b} ~]
{}~-~ 2 (\g^3)_{\a}{}^{\b} ( \nabla_{\b} {\rm R} ) {\cal M}  ~~~,
\eqno(A.8)$$
where ${\rm R} | \equiv {\rm B}$.  Since the effect in which we are
interested involves the bosonic fields, we will set all the fermionic
fields to zero. $$ {~~~}$$
$${\cal S} =
\int d^2 \s {\rm e}^{-1}  \{  ~ \ha \eta^{a b} ({\Hat \nabla}_a p ) (
{\Hat \nabla}_b p ) ~+~ \ha \eta^{a b} ({\Hat \nabla}_a \phi^i ) (
{\Hat \nabla}_b \phi_i ) ~+~ \ha (F_{,i})^2  {~~~~}  $$
$${~~~~~~~~~~~~~~~~~}  ~+~ F_{,i} {\rm U}_{,i} (\phi_i : p)
{}~-~ [{{\pa {~} \over \pa {p}}} {\rm U} (\phi_i : p) ]
(\e^{a b} F_{a b} (A) ~-~ p {\rm B} )$$
$${~~~~~~~~~~} ~+~\ha
 (\e^{a b} F_{a b} (A) ~-~ p {\rm B} )^2
-  {\rm B} {\rm U}(\phi_i : p) ~ \}
 ~~~, {~~~~~} \eqno(A.9)
$$
where
$$
\Phi_i | ~\equiv~ \phi_i  ~~~,~~~  P_i | ~\equiv~ p
{}~~~,~~~ \nabla^2 \Phi_i | ~\equiv~ F_i ~~~,
\eqno(A.10)$$
$$
{\Hat \nabla}_a ~=~ e_a {}^m \pa_m ~-~ \ha (\e^{b c} C_{b c}{}_a) {\cal M}
{}~+~ i g A_a t ~~~.
\eqno(A.11)
$$
Now eliminating the auxiliary fields by their algebraic equations of
motion,
$$ {\rm B} ~=~ p^{-1} \{~ \e^{a b} F_{a b} (A) ~-~ p^2 {{\pa {~}\over \pa {p}}}
[ p^{-1} {\rm U} (\phi_i : p) ~] ~\} ~~~,
\eqno(A.12)$$
$$ F_i ~=~ - {\rm U}_{,i} (\phi_i : p) ~~~,
\eqno(A.13)$$
and substituting these back into the action then yields
$${\cal S} =
\int d^2 \s {\rm e}^{-1}  \{  ~ \ha \eta^{a b} ({\Hat \nabla}_a p ) (
{\Hat \nabla}_b p ) ~+~ \ha \eta^{a b} ({\Hat \nabla}_a \phi^i ) (
{\Hat \nabla}_b \phi_i ) ~-~ \ha ({\rm U}_{,i} (\phi_i : p))^2  {~~~~~~}  $$
$$ {~~~~~~~~~~~~~~~~~~~~~~~~}
-  p^{-1}{\rm U} (\phi_i : p) ~\e^{a b} F_{a b} (A) ~-~\ha
 [p^2{{\pa {~}\over \pa {p}}} ( p^{-1} {\rm U} (\phi_i : p)) ~]^2  ~\}
 ~~~. {~} \eqno(A.14)
$$
The origin of why such massive cancellations occur is given in (A.4).
In any action where the field strength occurs via spinorial differentiation
of $P$, the removal of the supergravity auxiliary scalar field will
have this result.  The final result shows us that potentials that are
linear in $p$ have a preferred role. For such a potential all of the
second line of the action vanishes.  Additionally, the penultimate
2D-CS term, under these conditions, is independent of the ``prima-photon''
$p$ and the last term in the action also vanishes.  Finally, the most
unusual feature is that the spin-1 field strength does not appear at all
unless there is a non-trivial potential ${\rm U} (\Phi_i : P)$!  Since
all $N > 1$ theories must have an N = 1 theory embedded within them, this
proves our general assertion.

\newpage

\end{document}